\documentclass[prl,aps,letter,onecolumn]{revtex4}
\usepackage[french]{babel}
\usepackage{amsmath}
\usepackage{amssymb}
\usepackage[T1]{fontenc}
\usepackage{graphicx}

\begin{document}

\title{Photon-correlation \\Fourier spectroscopy}

\author{Xavier Brokmann, Moungi Bawendi}
\affiliation{Massachusetts Institute of Technology,
\\\mbox{77 Massachusetts Avenue, Cambridge, Massachusetts
02139.}}
\author{Laurent Coolen, Jean-Pierre Hermier}
\affiliation{Laboratoire Kastler Brossel,
\\\mbox{24, rue Lhomond, F-75005 Paris.}}

\begin{abstract}
We describe a method to probe the spectral fluctuations of a
transition over broad ranges of frequencies and timescales with the
high spectral resolution of Fourier spectroscopy, and a temporal
resolution as high as the excited state lifetime, even in the limit
of very low photocounting rates. The method derives from a simple
relation between the fluorescence spectral dynamics of a single
radiating dipole and its fluorescence intensity correlations at the
outputs of a continuously scanning Michelson interferometer. These
findings define an approach to investigate the fast fluorescence
spectral dynamics of single molecules and other faint light sources
beyond the time-resolution capabilities of standard spectroscopy
experiments.\\
$$ $$
\end{abstract}

\maketitle

%%%%%%%%%%%%%%%%%%%%%%%%%%  body  %%%%%%%%%%%%%%%%%%%%%%%%%%
\section{Introduction}

Chromophores embedded in a condensed medium inevitably exhibit
time-dependent, fluctuating optical properties reporting on the
dynamics of their nanoscale environment. Studied on ensembles of
molecules \cite{Haa90}, these fluctuations illuminate the complex
dynamics of a broad range of disordered host systems such as
low-temperature glasses, proteins and liquids, over timescales
extending from femtoseconds to hours \cite{Gev97,Fri98}. Observed
at the single emitter level, chromophore fluctuations reveal
surprisingly varied and complex dynamical phenomena kept hidden in
ensemble-averaged experiments \cite{Moe99}, whose understanding is
crucial for developing applications such as single molecule probes
in biophysics or single photon sources in quantum information
processing \cite{Wei99}.

Single molecule spectroscopy reaches its fullest potential when
combined with high time resolution, so as to resolve the fast
processes and temporal heterogeneities of any given isolated
emitter. Obtained down to timescales shorter than the excited
state lifetime for single emitter intensity fluctuations
\cite{Lip05}, high time resolution proves much more difficult to
achieve when probing spectral fluctuations, due to the finite
integration time (typically larger than milliseconds) necessary to
collect enough photons to measure the spectrum of the emitter
\cite{Boi99}.

The first attempt to achieve fast single molecule spectroscopy was
undertaken by Plakhotnik, who demonstrated that time resolution
could be improved by averaging the autocorrelation functions of
many individual fast scan spectra provided by single molecule
laser spectroscopy \cite{Pla98,Pla99}. The time resolution of this
intensity-time-frequency correlation (ITFC) method is however
technically limited to the order of milliseconds by the finite
scan rate of the laser, and no application followed this
pionneering work. As a result, our current understanding of fast
spectral fluctuations in single emitters is so far mostly inferred
from inhomogenously broadened ensemble experiments.

\begin{figure}[b]
\begin{center}
\includegraphics[width=7cm]{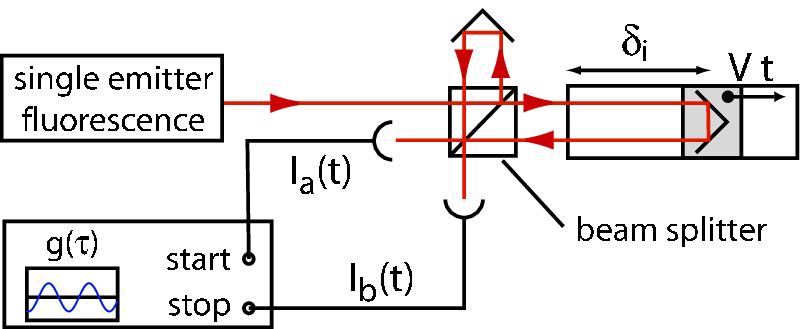}
\caption{Single molecule photon-correlation Fourier spectroscopy
setup. Starting from an initial optical path difference
$\delta_i$, the intensity correlation function $g(\tau)$ of the
output intensities $I_\mathrm{a}(t)$ and $I_\mathrm{b}(t)$ is
measured during a continuous scan of the interferometer at a
velocity~$V$. Repeating this procedure for various values of
$\delta_i$ provides the time-resolved frequency fluctuation
spectrum $p_\tau(\zeta)$ of the emitter.} \label{fig1}
\end{center}
\end{figure}

In this article, we describe an approach overcoming these
limitations to explore the spectral dynamics of a single
transition at both high temporal and high spectral resolutions
with standard photocounting equipment, and without requiring
mode-locked laser sources for excitation as in ITFC. The method is
based on the observation that spectral fluctuations under
continuous excitation can be directly encoded in the intensity
correlation functions measured at the output of a continuously
scanning Michelson interferometer by a pair of photodiodes
followed by a photon correlation counting board [Fig. \ref{fig1}].
The setup is investigated theoretically, simulated numerically,
and compared to standard spectroscopy experiments.

\section{Theoretical background}

\subsection{Intensity correlation functions at the output of a scanning interferometer}

The basis of our approach is the existence of an intimate relation
between the spectral fluctuations of a transition and the intensity
correlation function of its radiated field after transmission
through a scanning Michelson interferometer. We first derive this
relation in the case of a monochromatic transition undergoing random
stationnary fluctuations $\delta\omega(t)$ around a frequency
$\omega_0=\langle \omega(t)\rangle$ over a mean-squared range
$\sigma^2=\langle\delta \omega(t=0)^2\rangle$, and later generalize
to fluctuating transitions with any lineshape.

Starting from an initial optical path difference $\delta_i$ at
time $t=0$, the mirror travelling at a constant velocity $V$
creates a time dependent optical path difference
$\delta(t)=\delta_i+2Vt$ between the two arms of the
interferometer. After entering the interferometer, frequency
fluctuations propagate in the arms and recombine on the
beamsplitter. This recombination can be described as
instantaneous, as the retardation effects (of order $\delta_i/c$)
induced by optical path differences $\delta_i$ shorter than the
maximum possible coherence length $cT_1$ of the emitter remain
neglible over the timescales $\tau>T_1$ considered throughout this
paper (i.e. $\delta_i/c\ll\tau$. See also \cite{Coo06}). Hence, we
have :
\begin{equation}\label{iaib} \left\{\begin{array}{ccc}
I_\mathrm{a}(t) & \propto & 1+\cos[(2Vt+\delta_i)\omega(t)/c]\\
I_\mathrm{b}(t) & \propto & 1-\cos[(2Vt+\delta_i)\omega(t)/c].\\
\end{array}\right.
\end{equation}
These oscillating intensities feed a photon counting board
integrating photons between $t=0$ up to a time $t=T$ for computing
the time-averaged intensity correlation function between the two
outputs of the interferometer
\begin{equation}\label{defgab}
g(\tau)=\frac{\overline{I_\mathrm{a}(t)
I_\mathrm{b}(t+\tau)}}{\overline{I_\mathrm{a}(t)}\phantom{1}\overline{
I_\mathrm{b}(t+\tau)}},
\end{equation}
where $...\!\!\!\!\!\overline{\phantom{A}}$ denotes time-averaging
from $t=0$ up to $t=T$. The integration time $T$ is chosen so that
the corresponding change in optical path difference
$\Delta=\delta(T) -\delta(0)=2VT$ spans a large number of fringes,
reducing the time-averages in Equation \ref{defgab} to :
\begin{equation}\label{galpha}
g(\tau) = 1- \frac{1}{2T}\int_{0}^{T}\cos(2\omega_0
V\tau/c+\alpha(t)\delta_i/c)dt
\end{equation}
with $\alpha(t)=2\delta\omega(t+\tau)V\tau/\delta_i
+(1+2Vt/\delta_i)\zeta_{\tau}(t)$, where
$\zeta_{\tau}(t)=\delta\omega(t+\tau) -\delta\omega(t)$ denotes
the random frequency fluctuation observed between times $t$ and
$t+\tau$.

This general expression of $g(\tau)$ simplifies with an
appropriate choice of the scanning parameters $V$ and $\delta_i$,
since the first term of $\alpha(t)$ becomes negligible in
$g(\tau)$ at slow scanning velocities, when
$2\delta\omega(t+\tau)V\tau/\delta_i\sim 2\sigma V\tau/\delta_i\ll
c/\delta_i$, i.e. as the change in the optical path $2V\tau$
occuring over the timescale $\tau$ under investigation remains
small compared to the coherence length $\Lambda=c/\sigma$ of the
emitter. The quantity $\alpha(t)$ then reduces to
$\alpha(t)=\zeta_{\tau}(t)$ in Eq. \ref{galpha} when
$\delta_i\gg\Delta$, and the intensity correlation function
becomes :
$$g(\tau) = 1-\frac{1}{2T}\int_{0}^{T}\cos(2\omega_0 V\tau/c+
\zeta_{\tau}(t)\delta_i/c)dt.$$ For integration times $T$ much
larger than the typical timescale over which spectral fluctuations
occur, time averages can be replaced by the ensemble-average over
the distribution $p_{\tau}(\zeta)$ of all the possible
realizations of the random variable $\zeta=\zeta_{\tau}(t=0)$, so
$g(\tau)$ now becomes :
$$g(\tau) = 1
-\frac{1}{2}\int_{-\infty}^{+\infty}\cos(2\omega_0
V\tau/c+\zeta\delta_i/c)p_\tau(\zeta) d\zeta,$$ which can be recast
in a more compact form as :
\begin{equation}\label{leq}
g(\tau)=1-\frac{1}{2}\cos(2\omega_0
V\tau/c)\mathrm{FT}[p_\tau(\zeta)]_{\delta_i/c}
\end{equation} when the fluctuation process is time-reversal invariant, which
imposes $p_\tau(\zeta)=p_\tau(-\zeta)$.

Hence, provided the two conditions i) $2V\tau\ll c/\sigma$ and ii)
$\delta_i\gg\Delta$ are fulfilled, the time-averaged intensity
correlation function $g(\tau)$ measured at the output of a scanning
Michelson interferometer oscillates with the frequency $2\omega_0
V/c$ at which fringes oscillate on the photodetectors, and with an
amplitude given by the value of the Fourier transform $\mathrm{FT}
[p_\tau(\zeta)](\theta)$ at $\theta=\delta_i/c$, where
$p_\tau(\zeta)$ is the distribution of frequency shifts.

\subsection{General case : narrow transition with arbitrary lineshape }

The result obtained above holds for fluctuating transition of any
(narrow) lineshape, integration over the finite linewidth simply
changing the distribution of frequency shifts $p_\tau(\zeta)$ into
the more general expression :
\begin{equation}\label{pcon}
p_\tau(\zeta)=\langle\int_{-\infty}^{+\infty}s_t(\omega)
s_{t+\tau}(\omega+\zeta) d\omega\rangle,
\end{equation}
where $\langle...\rangle$ denotes ensemble averaging over all
possible realizations of spectral fluctuations, and $s_t(\omega)$
is the time-resolved emission spectrum of the transition
\cite{Gev97}~:
\begin{equation}\label{st}
s_t(\omega) = \frac{1}{\pi}\int_0^{+\infty} e^{-t'/2T_1}
\Re\big[e^{i\omega_0 t'}e^{i\int_0^{t'} {\delta
\omega(t+u)du}}\big]dt'
\end{equation}
where $T_1$ is the excited state lifetime of the transition.
Remarkably, Eq. \ref{pcon} identifies with the time-frequency
spectrum provided by past ITFC experiments, and is the Fourier
transform (in $\zeta$) of the echo obtained in three-pulse photon
echo spectroscopy \cite{Gev97,Pla99}.

\subsection{Photon-correlation Fourier spectroscopy}\label{pcfspart}

The key point is that at timescales $\tau$ shorter than the
periodicity $\pi c/\omega_0 V$ at which the fringes oscillate on the
photodetectors, condition i) (cf. 2.1) is automatically fulfilled
for any narrow spectral fluctuations (i.e. when $\sigma\ll\omega_0$)
and Eq. \ref{leq} can be reverted as :
\begin{equation}\label{leq2}
p_\tau(\zeta)=2\mathrm{FT}^{-1}[1-g(\tau)]_{\zeta=2\pi
c/\delta_i}.
\end{equation}
The distribution of the spectral fluctuations $p_\tau(\zeta)$ of the
emitter can therefore be determined directly from the measurement of
its fluorescence intensity correlation function $g(\tau)$ at various
optical path differences $\delta_i$ of the scanning interferometer.

Determining the distribution $p_\tau(\zeta)$ of an emitter rather
than its fluorescence spectrum $s_t(\omega)$ suggests an original
approach to investigate its fast spectral dynamics, very much in the
same way that measuring an intensity autocorrelation function
instead of an intensity time trace transformed the study of photon
statistics on short timescales in faint light sources such as
distant stars or nanoscale emitters \cite{HBT}. This can be
understood by taking a closer look at some of the general properties
of the distribution $p_\tau(\zeta)$. When $\tau\rightarrow\infty$,
the spectra $s_t(\omega)$ and $s_{t+\tau}(\omega)$ are statistically
independent, and so $p_\tau(\zeta)$ reduces to the autocorrelation
of the time-averaged (inhomogenous) linewidth $s(\omega-\omega_0)$
of the transition accessed by any standard fluorescence spectroscopy
experiment. On the contrary, in the limit $\tau\rightarrow0$,
$p_\tau(\zeta)$ is given by the autocorrelation of the time-resolved
(homogenous) Lorentzian lineshape of purely radiative width
$T_1^{-1}$, as no fluctuation has time to occur within delays
$\tau=0$. The distribution $p_\tau(\zeta)$ hence naturally bridges
the gap between single molecule fluorescence trajectory analysis and
ultra-fast spectroscopy ensemble experiments.

The time-frequency distribution $p_\tau(\zeta)$ also has the
property of being time-independent for any stationnary
distribution of fluctuations. The measurement of $p_\tau(\zeta)$
can therefore be made by integrating photon coincidences over long
durations to improve signal-to-noise ratio without degrading
temporal and spectral resolutions. Derived within the frame of
classical electrodynamics, our analysis extends down to timescales
$\tau_\mathrm{min}$ as short as the excited state lifetime of the
emitter $T_1$, readily accessible by the modified Hanbury-Brown
Twiss photon correlation detection setup described Fig. 1
\cite{Coo06}. This approach - which we call photon correlation
Fourier spectroscopy (PCFS) - therefore combines the high spectral
resolution $\zeta_\mathrm{min}\sim c/\delta_i$ of Fourier
spectroscopy with a high temporal resolution down to the
Fourier-transform limit $\tau_\mathrm{min} \zeta_\mathrm{min} \sim
1$, beyond the current capabilities of standard single molecule
experiments.

\section{Numerical simulations and discussion}\label{simulations}

Numerical simulations were performed to investigate the validity of
PCFS for exploring fast spectral fluctuations and highlight some of
its instrumental properties. All simulations were made assuming an
excited state lifetime $T_1=1$ ns, and fast random spectral
fluctuations over a total range $\delta\lambda=2\lambda_0 \sigma
/\omega_0=1$ nm around an average wavelength $\lambda_0=600$ nm,
with an exponential frequency correlation function $C(\tau)=
\langle\delta\omega(t) \delta\omega(t+\tau)\rangle$ of correlation
time $\tau_c=\int_0^\infty C(t)dt/\sigma^2=5$ $\mu$s. The choice of
these numerical parameters corresponds to typical values as
encountered in standard single molecule spectroscopy experiments for
emitters such as molecules or semiconductor quantum dots.

\begin{figure}
\begin{center}
\includegraphics[width=13cm]{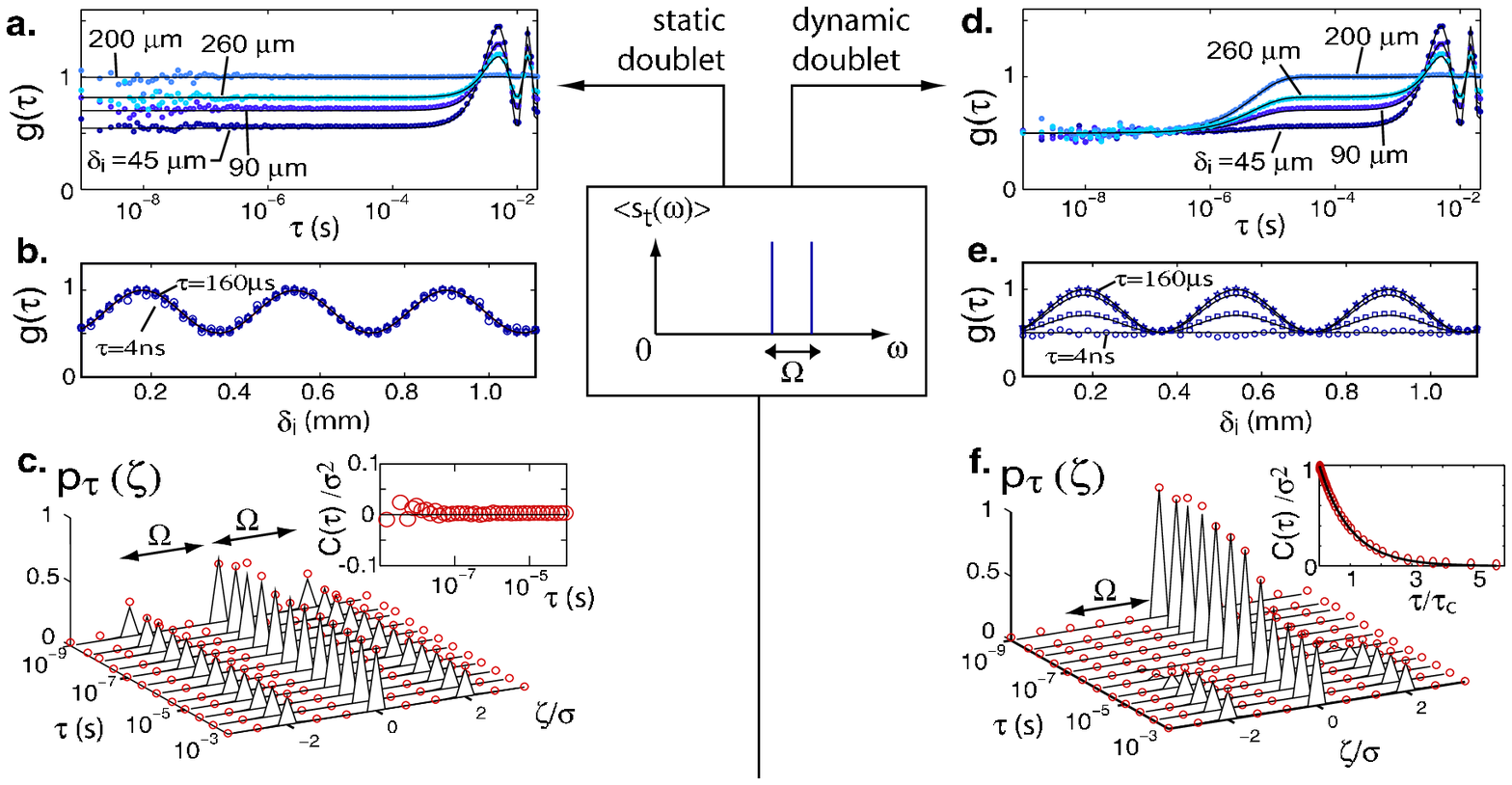}
\caption{Photon correlation spectroscopy of a single static (left)
or switching doublet (right). (a) Intensity correlation function
at various delays $\delta_i$. The scatter plots are numerical
simulations for an emitter detected with an intensity $I=50$ kHz.
(b) Evolution of $g(\tau)$ with $\delta_i$ for $\tau$=4 ns
($\circ$), 2.5 $\mu$s ($\square$), 10 $\mu$s ($\diamond$), 160
$\mu$s ($\star$), depending of the optical delay $\delta_i$ where
the measurement was performed. (c) Corresponding fluctuation
distribution $p_\tau(\zeta)$ ($\circ$). (d,e,f) Same as in (a,b,c)
for the switching doublet. Solid lines are the theoretical
expectations corresponding to the simulation parameters (see Table
1).} \label{fig2}
\end{center}
\end{figure}

PCFS measurements were numerically simulated with a speed of the
translation stage set to $V=30$ $\mu$m/s to fulfill \mbox{condition
i)} ($V\ll 1.8$ mm/s) over timescales $\tau<100$ ms. The photon
arrival times at the entrance of the interferometer were drawn
according to the Poissonian statistics of a light beam of intensity
$I=5\times 10^4$ photons/s. This corresponds to a situation where
spectral fluctuations occur at timescales $\tau_c=5$ $\mu$s much
faster than the average delay $I^{-1}=20$ $\mu$s between two
successive photodetection events, and so are completely averaged out
in standard single molecule spectroscopy experiments. The photons
were statistically directed towards either photodiode depending on
their wavelength and arrival time in the interferometer. Intensity
correlation functions were calculated for various $\delta_i$ by
integrating photons over 30 fringes over $500$ scans, corresponding
to a total acquisition time of 5 min per intensity correlation
function.

\begin{table}
\centering
\begin{tabular}{c c}

\hline\hline

Sudden jumps \parbox[c]{5cm}{\begin{flushright}
\includegraphics[width=3cm]{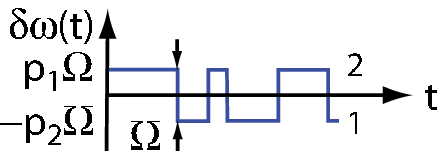}\end{flushright}} &
\parbox[c]{3.5cm}{\begin{flushleft} $\left|\begin{array}{l}\omega(t)=\omega_0 +\Omega S(t)\\
C(\tau) = \Omega^2\langle S(t)S(t+\tau)\rangle\\ \end{array}\right. $ \end{flushleft}} \\
\multicolumn{2}{c}{\parbox[c]{13cm} {\begin{center}
$g(\tau)=1-\frac{1}{2}\cos(2\omega_0 V\tau
/c)e^{-\delta_i/cT_1}[A_0+A_{\Omega}\cos(\Omega\delta_i/c)]$ \\
where
$\left\{\begin{array}{l}A_0=(p_1p_2)^2[p_1^2+p_2^2+2C(\tau)/\Omega^2]
\\ A_\Omega=2(p_1p_2)^2[C(0)-C(\tau)]/\Omega^2\end{array}\right.$ \end{center}}} \\

\hline \hline

Gaussian fluctuations \quad \parbox[c]{3.5cm}{\begin{flushright}
\includegraphics[width=3cm]{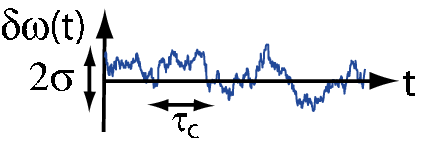}\end{flushright}} &
\parbox[c]{3.5cm}{\begin{flushleft}
$\left|\begin{array}{l} \mathrm{Prob}(\delta\omega) \propto
e^{-\delta\omega^2/2\sigma^2} \\ C(\tau)=\sigma^2e^{-\tau/\tau_c}\\ \end{array}\right. $ \end{flushleft}} \\
\multicolumn{2}{c}{\parbox[c]{8cm} {\begin{center}
$g(\tau)=1-\frac{1}{2}\cos(2\omega_0 V\tau
/c)e^{-\delta_i/cT_1}F_\tau(\delta_i/c)$ \\
where $F_\tau(t)=e^{-\int_0^t(t-t')[2C(t')
-C(t'+\tau)-C(t'-\tau)]dt'}$\end{center}}}\\

\hline\hline

\end{tabular}

\caption{\label{tab} Theoretical expression of the intensity
correlation function $g(\tau)$ measured in PCFS for discrete and
continuous spectral fluctuations. $C(\tau)=
\langle\delta\omega(t)\delta\omega(t+\tau)\rangle$ is the
frequency correlation function of the fluctuations. $p_{i=1,2}$
denote the fraction of time spent by the transition in states 1
and 2 respectively.}
\end{table}

\subsection{PCFS and spectral fluctuation dynamics analysis}

We first investigated the outcome of a PCFS experiment for a
transition undergoing discrete spectral fluctuations, with a
frequency $\omega(t)$ switching between two values $\omega_1$ and
$\omega_2=\omega_1+\Omega$ (here corresponding to wavelength jumps
of $\pm1$ nm) as a random telegraph signal $S(t)$ [Fig. 2, Tab. 1]
as encountered in the study of chromophores interacting with the
two-level systems in glasses at low temperature~\cite{Gev97}. The
transition frequency fluctuations were generated with
exponentially distributed waiting times of rate $k_1$ and $k_2$
with $k_1=k_2$ (and $\tau_c=[k_1+k_2]^{-1}$). Observed in standard
spectroscopy, the transition spectrum would appear as predicted by
the Anderson-Kubo lineshape theory \cite{And54}, i.e. a doublet of
separation $\Omega=2\sigma$ centered in $[\omega_1+\omega_2]/2$,
indiscernible from the spectrum of a static doublet of transitions
at frequencies $\omega_1$ and $\omega_2$ with similar intensities.

Figure 2 shows the result of PCFS experiments on a static doublet
(left), and on the switching transition (right). As seen in Fig.
2(a)(d), in both cases, the shape of the intensity correlation
function strongly depends of the delay $\delta_i$ where it was
measured. Repeating the measurement of $g(\tau)$ over different
optical path differences $\delta_i$, we determined the evolution
of $g(\tau)$ with $\delta_i$ [Fig. 2(b)(e)], from which the
distribution $p_\tau(\zeta)$ was extracted with Eq. \ref{leq2}
[Fig. 2(c)(f)]. For a static spectrum, photons do not exhibit
spectral correlation, and so $p_\tau(\zeta)$ reduces to the
autocorrelation of the time-averaged spectrum measured in standard
spectroscopy (here a doublet), i.e. a triplet of lines of
intensities \{1/4,1/2,1/4\} at frequencies $\{-\Omega,0,+\Omega\}$
independently of the timescale $\tau$, as observed in the
experimental results shown Fig. 2(c).

For the switching transition, this pattern is only preserved over
timescales where fluctuations are uncorrelated, i.e. when
$\tau>\tau_c=5$ $\mu$s [Fig. 2(f)], and breaks down as soon as
$\tau\sim\tau_c$, when the sidebands of $p_\tau(\zeta)$ in
$\zeta=\pm\Omega$ decay progressively as $\tau$ decreases,
asymptotically leaving us with the autocorrelation of the time
resolved spectrum of the transition as $\tau\rightarrow0$ - here a
Lorentzian of width $T_1^{-1}$. Interestingly, the calculation of
$g(\tau)$ from Eq. \ref{leq}-\ref{st} (cf. Tab. 1) indicates that
the correlation function of the random telegraph signal governing
the frequency switching process is directly encoded in the decay of
the sidebands amplitude $A_\Omega$ with $\tau$.

The insets in Fig. 2(c)(f) - showing the correlation functions
$C(\tau)$ extracted from the decay of the measured sideband
amplitudes $A_\Omega$ - confirm this prediction. Indeed, they
indicate respectively that the static and dynamic doublet exhibit
null and exponentially decaying correlation function, as expected.
Illustrated here on binary spectral jumps, the ability of PCFS to
investigate spectral fluctuations dynamics down to the excited state
lifetime of the emitter can also be exploited on a transition
coupled to a collection of flipping two-level systems, for example
to determine the detailed physical properties of the latter (energy
splittings $\Omega$, switching timescales $\tau_c$, etc.)
\cite{Pla99}.

\subsection{High-resolution spectroscopy beyond temporal inhomogenous broadening}

PCFS also opens a perspective for performing high resolution
spectroscopy despite the presence of broad, fast, continuous
spectral fluctuations, as reported on most emitters - often down
to low temperatures (4.2 K) \cite{Orr02}. This point is
illustrated Fig. 3, showing the result of a PCFS experiment
simulated for a doublet undergoing fast stationnary Gaussian
fluctuations over a range $\sigma$ much broader than the doublet
separation $\Omega$ (here $\sigma=5\Omega$), with an exponential
correlation function $C(\tau)$ (Ornstein-Uhlenbeck fluctuation
process) with $\tau_c=5$ $\mu$s.

\begin{figure}[t]
\begin{center}
\includegraphics[width=6cm]{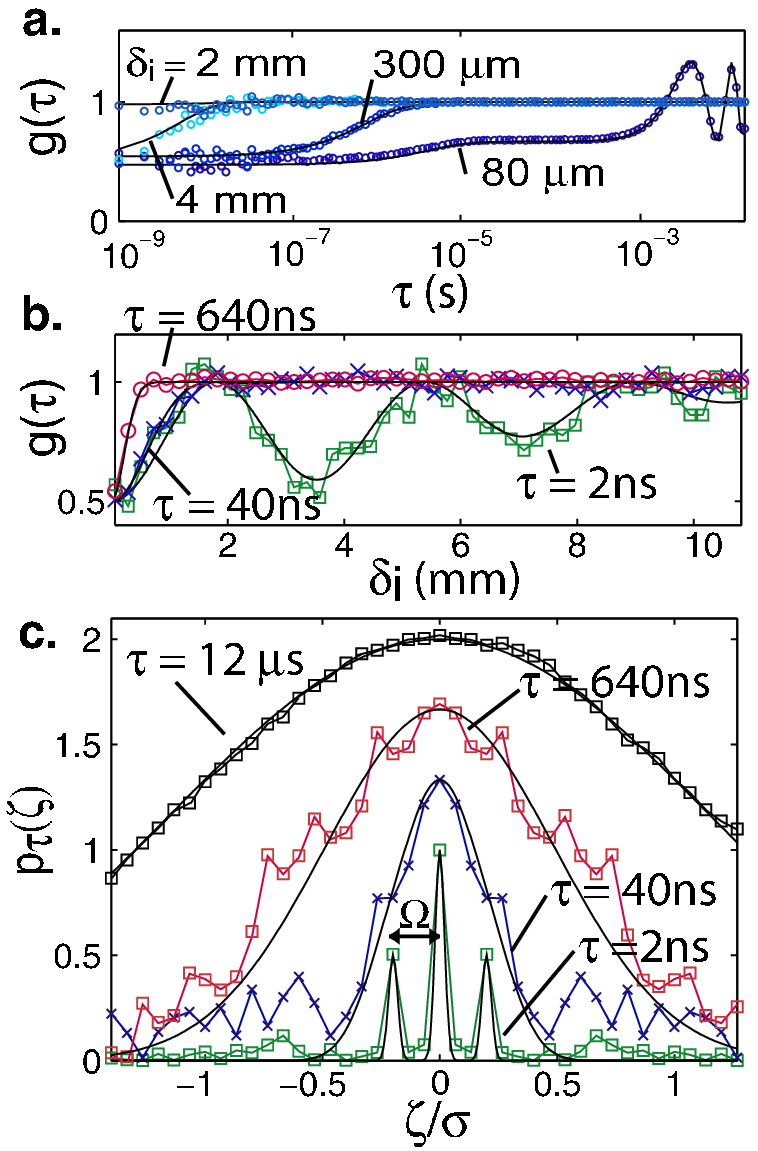}
\caption{Photon correlation spectroscopy of a doublet of
separation $\Omega$ undergoing Gaussian stationnary fluctuations
of correlation time $\tau_c=5\mu$s, over a spectral range
$\sigma=5\Omega$ (corresponding to $\delta\lambda=1$nm). (a)
Intensity correlation function at various delays $\delta_i$
obtained from numerical simulations when the emitter is detected
with an intensity $I=50$ kHz. (b) Evolution of $g(\tau)$ with
$\delta_i$ for $\tau$=2 ns ($\square$), 40 ns ($\times$), 640 ns
($\circ$) as observed from the measurement of $g(\tau)$, depending
of the optical delay $\delta_i$ where the measurement was made. At
short timescales ($\tau<10$ ns), oscillations of periodicity $2\pi
c/\Omega$ are observed, as the doublet becomes resolved. (c)
Corresponding fluctuation distribution $p_\tau(\zeta)$. At short
timescales, a triplet appear, i.e. the doublet is resolved. Solid
lines are the theoretical expectations corresponding to the
simulation parameters (see Table 1).}\label{fig3}
\end{center}
\end{figure}

Observed in conventional spectroscopy, the transition would appear
as a Gaussian lineshape of width $\sigma$, and the existence of the
underlying doublet would remain unnoticed. PCFS, in comparison,
reveals a completely different pattern. Here again, the intensity
correlation functions were found strongly dependent of $\delta_i$
[Fig. 3(a)], and, measured for different values of $\delta_i$,
provided the distribution $p_\tau(\zeta)$ [Fig. 3(b)(c)]. As seen in
Fig. 3(c), the distribution $p_\tau(\zeta)$ - broad (here a Gaussian
of FWHM=$2\sigma$) over durations $\tau>\tau_c$ - progressively
narrows when $\tau<\tau_c$. This is consistent with the fact that
for photons separated by delays $\tau$ shorter than the fluctuation
correlation time $\tau_c$, fluctuations are increasingly seen as
``frozen'' as $\tau\rightarrow 0$.

Due to this fluorescence line-narrowing effect, the doublet can be
resolved over short timescales, as seen from the quasi-periodic
oscillations (of periodicity $2\pi c/\Omega$ in $\delta_i$)
appearing in the intensity correlation function $g(\tau)$ when
$\tau\sim 5$ ns [Fig. 3(b)], which translate into the lineshape of
$p_{\tau<5\mathrm{ns}}(\zeta)$ expected for a static doublet, i.e.
a triplet of intensities (1/4,1/2,1/4) at frequencies
$(-\Omega,0,\Omega)$ [Fig. 3(c)]. A corollary of the fluorescence
line-narrowing effect is an increased coherence length of the
emitter, coherence in the transition emission (i.e. $g(\tau)<1$ in
[Fig. 3(b)]) remaining visible over short timescales
$\tau\ll\tau_c$ although we have $\delta_i>5$ mm $\gg\Lambda$,
i.e. even if the optical path difference is much larger than the
bare coherence length $\Lambda=120$ $\mu$m of the emitter, when no
fringe would be observed in standard Fourier spectroscopy.

\section{Experimental properties}

From a technical standpoint, the intensity correlation
measurements at the basis of PCFS can be considered as an
intensity homodyne detection, where the oscillating intensity
detected by a photodiode is demodulated by the oscillating
intensity detected on the other photodiode. A first consequence of
this observation is that PCFS directly provides the envelope of
the Fourier transform interferogram of the radiation field,
without the complex demodulation schemes usually involved in
scanning Fourier spectroscopy.

Secondly, the self-demodulation process implies that the shape of
$g(\tau)$ is robust against fluctuations in the scanning velocity
$V$ (as caused by stick-slip and vibrations in the translation
stage), and is independent of the exact average frequency $\omega_0$
of the transition, making PCFS intrinsically insensitive to rare,
large spectral jumps which often limit the measurement time of laser
spectroscopy experiments (e.g. the PCFS inversion formula (Eq. 8) is
indeed independent of $V$ and $\omega_0$). Numerical simulations
(not shown) confirmed this analysis. For example, random velocity
fluctuations of 30\% were found to have no significant impact on any
of the simulated results presented in Fig. 2 and Fig. 3.

Finally, we note that PCFS also offers high time and high spectral
resolution over a broad range of frequencies and timescales,
contrasting with laser spectroscopy, which - because of its very
scanning nature - only provides high time and spectral resolutions
simultaneously in the limit of vanishing spectral ranges.

\section{Conclusion}

Replacing the beamsplitter of a Hanbury-Brown Twiss detection
system by a scanning Michelson interferometer allows the
measurement of spectral fluctuations of a transition at high
spectral resolution, down to timescales as short as the transition
excited state lifetime, which opens unexplored possibilities for
studying the fast fluorescence spectral dynamics of single dipoles
in a range of contexts.

In solid state physics, this photon-correlation spectroscopy
method suggests an approach to investigate the relation between
lineshape broadening and decoherence in molecules and
semiconductor quantum dots, as well as the dynamical interactions
of these systems with optical fields and their nanoscale
environment. Implemented in a fluorescence correlation
spectroscopy (FCS) experiment, PCFS might also provide some
insight into the spectral dynamics of nanoscale emitters under the
influence of chemical reactions, conformational changes or
intermolecular interactions in liquid environments.

\section*{Acknowledgments}

We are grateful to J. Enderlein for his help in the fast computation
of intensity correlation functions \cite{End03}. This research was
funded in part through the NSF-Materials Research Science and
Engineering Center Program (DMR-0213282) and the Packard Foundation.

%%%%%%%%%%%%%%%%%%%%%%% References %%%%%%%%%%%%%%%%%%%%%%%%%


\begin{thebibliography}{99}
\bibitem{Coo06}{The extension of this method to delays $\tau<T_1$ pertains
to quantum electrodynamics - so as to account for photon
antibunching and photon coalescence effects - and will be analyzed
in detail in a forthcoming paper entitled ``Spectral diffusion and
time-coherence of single photons''.}
\end{thebibliography}
\end{document}